\documentclass[twocolumn]{article}
\input{epsf.sty}
\usepackage{epsfig}
\usepackage{multicol}
\usepackage{array}
\usepackage{amsfonts,amsmath,amssymb}

\begin{document}

\begin{center}
{\Large {\bf Resonance line in rotating accretion disc}}
\medskip
\end{center}
\begin{center}
{N. A. Silant'ev\thanks{E-mail: nsilant@bk.ru}\,\,, G. A. Alekseeva,\, Yu. K. Ananjevskaja,\,V. V. Novikov}
\medskip\\
\end{center}
\begin{center}
{ Central Astronomical Observatory at Pulkovo of Russian Academy of Sciences,\\ 196140,
Saint-Petersburg, Pulkovskoe shosse 65, Russia}
\medskip
\end{center}
\medskip
\begin{center}
{received..... 2018, \qquad accepted....}
\end{center}
\begin{abstract}

We  study the resonance  line emission from the rotating plane optically thick accretion disc, consisting of  free electrons and  resonant atoms.  We use the standard assumption that the source of continuum radiation is located near central plane of the accretion disc, where the temperature is the highest. This corresponds to the Milne problem consideration for continuum.    We shortly discuss the impossibility of the Milne problem for the resonance radiation. We  assume that the resonant atoms are located in a thin layer of an accretion disc near the surface.  In this case the resonance line emission arises due to  scattering of a continuum on the resonant atoms.  In   thin layer we can neglect the multiple scattering of the resonance radiation on  the  resonant atoms.  We consider the axially symmetric problems, where the Stokes parameter U =0.
 
 We take into account the Doppler effect for the frequencies of the resonance line. The three types of the resonant atom sources are considered  (see Figs.1-3). The first source is the axially symmetric continuous distribution of the resonant atoms along  the circular orbit. The second spot-like  source  rotates in the orbit. The third type presents two spot-like sources located in the orbit contrary one to another. In the first and third  cases the shape of the emitting resonance line is symmetric, i.e. the right and left wings have the similar shapes. In the second case the resonance  line has asymmetric shape.  The shape of the emerging line depends significantly on the  ratio of the rotation velocity value to the velocity, characterizing the Doppler width. It also  depends on the ratio of the electron number density to the number density of resonant atoms.   The results of the calculations characterize the different  observational effects  of H$\alpha$ radiation in the accretion discs and can be used for estimations of the parameters mentioned above. They also can be used for estimation of the inclination angle of an accretion disc. 
 \end{abstract}


$^1$ E-mail: nsilant@bk.ru

\section{Introduction}
 The observations of resonant lines emission  gives many information about the radiating atmospheres. The observation of the spectral line polarization  increases the number of an information. As an example, we consider the polarized line emission from Seyfert galaxies  presented in a number of the papers (see, for example, Antonucci 1984 ; Antonucci \& Miller 1985 ; Smith et al. 2002, 2004, 2005, Marin 2014). 

 There is  the standard explanation of the differences between Seyfert-1 and Seyfert-2 galaxies. These differences arise from the  orientation effect and the  presence of gas-dusty torus in an equatorial plane (see Antonucci 1993). According to this explanation, the line of sight ${\bf n}$ to Seyfert-1 galaxies lies near the normal ${\bf N}$ to the equatorial plane whereas the Seyfert-2 are observed near the direction to gas-dusty torus. The position angle  of polarization (the wave electric field oscillations) in the first case lies in the plane ${\bf (nN)}$ and in the case of Seyfert-2 galaxies the direction of the wave electric oscillations is perpendicular to the plane ${\bf (nN)}$. Of course, there are numerous exceptions  (see Goodrich \& Miller 1994, Martel 1996, Corbett et al. 1998).

 There are many theoretical papers devoted to calculation of the intensity and  polarization of the resonance  line emission ( for example, Ivanov 1963, 1973, 1995; Nagirner 1964, Nagirner \& Ivanov 1966; Faurobert 1988, Faurobert \& Frish, 1989, Faurobert, Frish \& Nagendra 1997; Fluri 2003, Dementyev 2008, Silant'ev et al. 2017a.) They do not consider the influence of the source rotation on the shapes of the resonance  line intensity and  polarization.

It should be noted that there are different assumptions about  the
places, where the emission of the  resonance line holds.  We consider the cases when the sources of  the resonant atoms (H$\alpha$) exist on the surface of an optically thick rotating  accretion disc. The rotation velocity mainly is due to the Keplerian rotation. The existence of the viscosity and the magnetic field gives rise to the slow fall of the substance to the center of a disc. In our calculations we neglect of this fall, considering the circle rotation. To explain the observed various shapes of the intensity and polarization degree, we consider three types of 
the resonant atoms  sources on the surface of the rotating accretion discs (see Figs. 1,2,3).

\begin{figure}
\includegraphics[width=1\columnwidth]{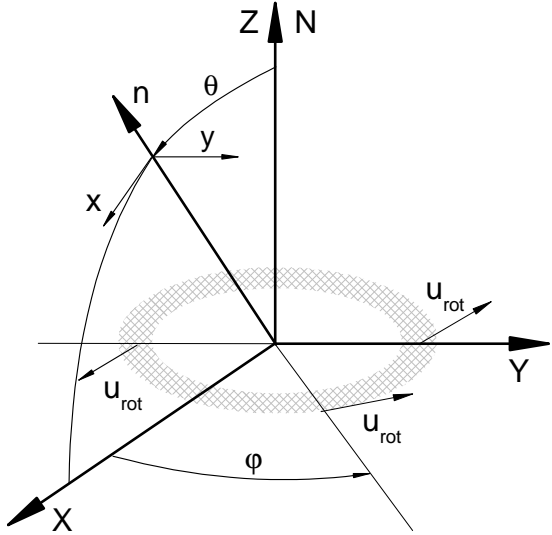}
\caption{Continuous distribution of resonant atoms.}       
\label{Silan_Fig1.eps}
\end{figure}

  The first source is the axially symmetric continuous distribution of the resonant atoms along  the circular orbit (see Fig.1). The second type corresponds to the small (point-like) source, located on the orbital place, where the Doppler change of the line frequency is maximum (see Fig.2).  The third type corresponds to two point-like sources located on the orbit  contrary one to another (see Fig.3). 

  Thus, we  study the influence of the accretion disc rotation on the intensity and polarization shapes  of a resonance line emission. The results of our numerical calculations we 
compare with the observed spectra of H$\alpha$ in Smith et al. 2002.

 Our Tables 1-3 and Figs. 4 - 9 allow to estimate the numerical values of the different parameters ( the ratio of the rotation velocity $u_{rot}$ to the velocity $u_0$, which  determine the Doppler shape of a line;  the ratio of the electron number density $N_e$ to the number density $N_{res}$ of the resonant atoms ,  and the inclination angle $i$ of the accretion disc plane).  These parameters are the basic ones for the construction of the source models.  Apparently, to estimate these parameters  the approximation procedure can be frequently used.
 
\begin{figure}
\includegraphics[width=1\columnwidth]{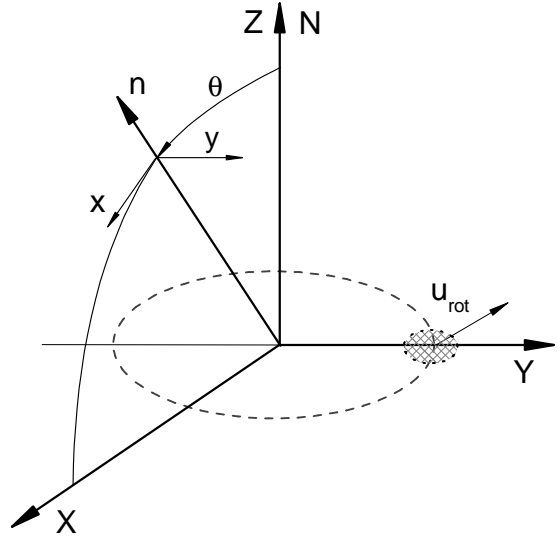}
\caption{Point-like distribution of resonant at $\varphi=90^{\circ}$.}       
\label{Silan_Fig2.eps}
\end{figure}

 The different problems (exponential, homogeneous and linearly increased sources of the resonant atoms)  for the small parameter $\beta\sim N_e/N_{res}$ were solved in Silant'ev et al. 2017a. In this paper the accretion disc is assumed to be  non-rotating.

\begin{figure}
\includegraphics[width=1\columnwidth]{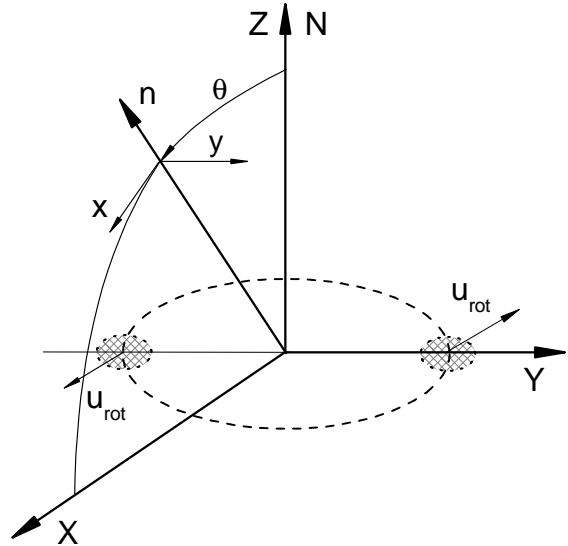}
\caption{Two point-like distribution of resonant atoms at $\varphi=90^{\circ}, 270^{\circ}$.}       
\label{Silan_Fig3.eps}
\end{figure}

 Below we consider the resonance line emission  from the optically thick rotating accretion disc with the arbitrary value of parameter $\beta$ . We assume that the resonant atoms are located in a thin layer below the surface. The source of continuum radiation is located near the central plane of an accretion disc, where the temperature is the highest and the intensive thermal non-polarized radiation arises (see Shakura \& Sunyaev 1973).  The continuum radiation near the accretion disc surface corresponds to the solution of the Milne problem.

 Recall, that the traditional Milne problem  describes the propagation of
radiation through the conservative (without absorption) optically thick atmosphere,  when the source of thermal radiation located far
below the surface. In this  case the  flux of the propagating continuum radiation is the same at every vertical distance in an atmosphere.

  The generalized Milne problem describes the propagation of continuum radiation in 
 the  non-conservative plane-parallel optically thick atmosphere. Due to existence of the absorption,  the radiation flux diminishes with the increasing of vertical distance from the cental accretion plane. The solutions of  both Milne problems give the angular distribution and polarization degree of emerging radiation (see Silant'ev et al. 2017b).

The goal of our paper is to calcutate the resonance radiation emission from an rotating  accretion disc for different types of the resonant atom sources.  We consider the  scattering of the continuum radiation  on the resonant atoms  near the surface of the  accretion disc as a mechanism  of the resonance emission.

 Note that the Stokes parameter $Q=I_x - I_y$, where $ I_x$ and $I_y$ are the radiation intensities with the wave electric field oscillations along the axes $x$ and $y$, correspondingly. The $x$-axis lies in the plane $({\bf nN})$ and the $y$ one is perpendicular to this plane. Here ${\bf N}$ is the normal to the plane-parallel atmosphere and ${\bf n}$ is the direction of the light propagation (see Fig.1). The X,Y,Z -  directions characterize the general reference frame in the accretion disc.

\section{Basic equations}
The radiative transfer equation for the vector (column) consisting of intensity $I_{res}(z,x,\mu)$ and the Stokes parameter $Q_{res}(z,x,\mu)$  of the spectral line emission has the form:

\[
\mu\frac{d{\bf I}_{res}(z,x,\mu)}{dz}=-[N_{res}\sigma^{(t)}_{res}\varphi(x)+N_e\sigma_T]{\bf I}_{res}(z,x,\mu)
\]
\[
+N_{res}\sigma^{(t)}_{res}\frac{(1-\varepsilon)}{2}\varphi(x)\hat A(\mu)\times
\]
\begin{equation}
\left[\int\limits_{-1}^1\,d\mu\int\limits_{-\infty}^{\infty}dx\,\varphi(x')\hat A^T(\mu){\bf I}_{res}(z,x',\mu)+\left (\begin{array}{c}s_I(z)\\s_Q(z) \end{array}\right)\right]
\label{1}
\end{equation}
\noindent  Here $\mu=\cos\theta$ with $\theta$ being the angle between the line of sight ${\bf n}$  and the normal ${\bf N}$ to the plane  optically thick  accretion disc; the value x is the dimensionless frequency (see Eq.(4)). The optical depth along the normal ${\bf N}$ is $d\tau=d\tau_{res}=N_{res}\sigma^{(t)}_{res}dz$, where $N_{res}$ is the number density of the resonant atoms and $\sigma^{(t)}_{res}$ is the frequency averaged total cross-section of the resonance radiation. The value  $N_e$ is the number density of free electrons, $\sigma_T$ is the Thomson cross-section.
The factor $(1-\varepsilon)=\sigma^{(s)}_{res}/\sigma^{(t)}_{res}$ is the probability of the resonance line emission at the scattering event.

 The dimensionless absorption factor is equal to $\alpha(x)=\varphi(x)+\beta$, where $\varphi(x)$  characterises the shape of
emission line. In general case the parameter  $\beta$ describes the extinction  of the  resonance line by all the factors  (free electrons, non-resonant atoms, the dust grains.) The extinction by free electrons is describes in Eq.(5).  Note that Eq.(1) does not describe the Compton scattering.  We consider only elastic scattering.  The values $s_I(z)$ and $s_Q(z)$ describe the sources for parameters $I_{res}(z,x,\mu)$ and  $Q_{res}(z,x,\mu)$ . The  values $I_{res}(z,x,\mu)$ and  $Q_{res}(z,x,\mu)$  have the dimension  $[\,erg/cm^2\, sec\, Hz  \,sterad\,]$.  The source term with $\hat{A}(\mu)\varphi(x){\bf s}(z)$ describes the scattered  radiation.

 Introducing the optical depth  $d\tau=N_{res}\sigma^{(t)}_{res}dz$, we obtain the standard form of the transfer equation  (see, for example, Ivanov 1995, Dementyev 2008, Silant'ev et al. 2017a):

\begin{equation}
\mu\frac{d{\bf I}_{res}(\tau,x,\mu)}{d\tau}=-\alpha(x){\bf I}_{res}(\tau,x,\mu)+\varphi(x)\hat A(\mu){\bf S}(\tau), 
\label{2}
\end{equation}
 where the vector ${\bf S}(\tau)$ is:

\[
{\bf S}(\tau)=\frac{1-\varepsilon}{2}\left (\begin{array}{c}s_I(\tau)\\s_Q(\tau) \end{array}\right) + {\bf K}_{res}(\tau),
\]
\begin{equation}
{\bf K}_{res}(\tau)=  \frac{1-\varepsilon}{2}\int\limits_{-1}^1\,d\mu\int\limits_{-\infty}^{\infty}dx\,\varphi(x)\hat A^T(\mu){\bf I}_{res}(\tau,x,\mu).
\label{3}
\end{equation}
\noindent The values  ${\bf S}(\tau)$ and  ${\bf K}(\tau)$ have the sense of the radiation density $[\,erg/cm^3\, sec\, Hz  \,sterad\,]$.

 Note that radiative transfer equation for the continuum radiation  ${\bf I}_{cont}(\tau,\mu)$  in the free electron atmosphere has the similar  form, if we omit the integration over frequencies and take $\varphi(x)=1$, $\beta=0$,  $\alpha(x)=1$, and $d\tau=N_e\sigma_Tdz$. The parameter $\varepsilon$ in this case denotes the probability of absorption  in the scattering act (see Eqs.(33, 34)).

 Usually one considers the multiple scattering of continuum radiation on free electrons only, taking into account that the Thomson scattering  cross-section  $\sigma_T\simeq 6.5 \cdot10^{-25}$ cm$^2$ 
is much greater than that  for atoms  $\sigma_{atom}\sim 10^{-28}$cm$^2$ . Thus, the scattering on non-resonant atoms can be neglected.

We take the Doppler shape of a  resonance line:
 
\begin{equation}
\varphi(x)=\frac{1}{\sqrt{\pi }}\exp{(-x^2)}.
\label{4}
\end{equation}
\noindent Dimensionless frequency is equal to $x=(\nu -\nu_0)/\Delta\nu_D$, where $\nu_0$ is the frequency in the center of a line. The Doppler width $\Delta\nu_D\equiv (\nu_0/c)u_0$.  The characteristic velocity  $u_0$ is equal to: $u_0= \sqrt{u_{th}^2+u_{turb}^2/3}$, where the thermal velocity along the line of sight is $u_{th}=\sqrt{2kT/M}$, and the turbulent velocity is the mean velocity of chaotic motions $u^2_{turb}=\langle {\bf u}({\bf  r},t)^2\rangle$ , $c$ is the radiation speed.

 The value $\Delta\nu=\pm(\nu-\nu_0)$ characterizes the shape of the line. It connected with $x$ by the formula   $\Delta\nu=\pm x\,\nu_0 (u_0/c)$ ($u_0/c << 1$). This means that the $\nu$-shape of the line looks much narrower than the $x$-shape. To transform the theoretical $x$-shape to the observed $\nu$-shape we  have to  know or assume the $u_0$-parameter.

The factor $\beta$ in the electron atmosphere is:
\begin{equation}
\beta=\frac{\tau_T}{\tau_{res}}=\frac{N_e\sigma_T}{N_{res}\sigma^{(t)}_{res}}\sim 10^{-9}\frac{N_e}{N_{res}},
\label{5}
\end{equation}
\noindent  where we take into account that the ratio $\sigma _T/\sigma^{(t)}_{res}\sim 10^{-9}$ .  Parameter $(1-\varepsilon)=\sigma^{(s)}_{res}/\sigma^{(t)}_{res}$ is the probability of the radiation scattering on a resonant atom. Parameter $\varepsilon=\sigma^{(a)}_{res}/\sigma^{(t)}_{res}\simeq 10^{-3}- 10^{-4}$ is the destruction probability (see Frisch \&  Frisch 1977). This parameter arises due to the two-photon radiation from the radiating atom level.  Note that this  parameter does not equal to zero.

The matrix $\hat A(\mu)$ is :
\begin{equation}
\hat A(\mu)= \left (\begin{array}{rr}1 ,\,\, \sqrt{\frac{W}{8}}(1-3\mu^2) \\ 0 , \,\,3\sqrt{\frac{W}{8}}(1-\mu^2) \end{array}\right).
\label{6}
\end{equation}
\noindent   The value $\hat A(\mu)\hat A^T(\mu) $ is the scattering matrix.  The parameter $W$ depends on the quantum numbers of the  transition atomic levels. For the simplest case of the dipole transition $W=1$.  
Our calculations  correspond to this case.   The superscript T is used for the matrix transpose.

 Using the formal solution of Eq.(2) (see Chandrasekhar 1960, Silant'ev et al. 2015), we derive the integral equation
 for  ${\bf S}(\tau)$:
\begin{equation}
{\bf S}(\tau)={\bf g}(\tau)+
\int\limits_0^{\infty} d\tau'\hat L(|\tau-\tau'|)\,{\bf S}(\tau').
\label{7}
\end{equation}
\noindent  The  term ${\bf g}(\tau)$ has the form:
\begin{equation}
{\bf g}(\tau)= \frac{1-\varepsilon}{2} \left (\begin{array}{c}s_I(\tau)\\s_Q(\tau) \end{array}\right).
\label{8}
\end{equation}
 The matrix kernel of  integral equation (7) is equal to:
\[
\hat L(|\tau-\tau'|)=\int\limits_0^1\frac{d\mu}{\mu}\int\limits_{-\infty}^{\infty}dx\,\varphi^2(x)\times
\]
\begin{equation}
\exp{\left(-\frac{\alpha(x)|\tau-\tau'|}{\mu}\right)}\,\hat \Psi(\mu),
\label{9}
\end{equation}
\noindent where
\begin{equation}
\hat \Psi(\mu)=\frac{1-\varepsilon}{2}\hat A(\mu)^T\hat A(\mu).
\label{10}
\end{equation}
\noindent   The matrix $\hat \Psi$, having the sense of the angular scattering function,  is symmetric: $\hat \Psi^T=\hat \Psi$. This property gives rise to the symmetry of the kernel $\hat L^T=\hat L$. The explicit form of matrix $\hat \Psi(\mu)$ is the following:
\[
\hat\Psi(\mu)=
\]
\begin{equation}
\frac{1-\varepsilon}{2} \left (\begin{array}{rr}1,\qquad \,\,\,  \sqrt{\frac{W}{8}}(1-3\mu^2) \\ \sqrt{\frac {W}{8}}(1-3\mu^2), \,\, \frac{W}{4}(5-12\mu^2+9\mu^4) \end{array}\right).
\label{11}
\end{equation}

Note that the resonance line shape (4) corresponds to the line frequency near the center of the line. Far from the center the resonance line shape is described by the Lorentz formula:
\begin{equation}
\varphi_{L}(x)=\frac{1}{\pi}\cdot \frac{\delta}{\delta^2+x^2}\,,
\label{12}
\end{equation}
\noindent where parameter $\delta=\Delta_{\gamma}/\Delta_{D}$ is the ratio of the Lorentz width to the Doppler one. The Lorentz width $\Delta_{\gamma}$ is proportional to the natural width of the line. Usully this width is much smaller than the Doppler width, i.e. the parameter $\delta<< 1$.

 It is of interest to know the frequency $x_{*}$ when $\varphi(x_{*})=\varphi_{L}(x_{*})$. According to Ivanov 1973,  for $\delta=0.01$ we have the value $x_{*}=2.67$ , for $\delta=0.001$ the value $x_{*}=3.12$ and for $\delta=0.0001$ the value $x_{*}=3.51$.
It appears that the results with the Doppler shape function (4) are valid for $x\le x_{*}$.

In the rotating  plane accretion disc  the Doppler effect gives rise to the following change of the frequency $\nu$:
\begin{equation}
\nu\to \nu- \frac{\nu_0}{c}u_{rot}\sin\theta\sin\varphi.
\label{13}
\end{equation}
\noindent Here we assumed the right-hand rotation with the azimuthal velocity $u_{rot}$. The azimuthal angle of rotating atoms $\varphi$ is assumed  to be zero for the $X$-axis. The line of sight ${\bf n}$ lies in the plane $({\bf X}{\bf N})$,  where ${\bf X}$, ${\bf Z=N}$ and ${\bf Y}$ are the corresponding unit vectors of the coordinate system. Recall, that the angle $\theta$  is the angle between the line of sight ${\bf n}$ and the  vector ${\bf N}$ (see Fig.1). This angle  is the inclination angle of an accretion disc, frequently denoted as $i$.

In the rotating accretion disc the shape (4) of the emission line transforms to:
\begin{equation}
\varphi(x)=\frac{1}{\sqrt{\pi }}\exp{\left[-\left(x-\frac{u_{rot}}{u_0}\sin\theta\sin\varphi \right)^2\right]}.
\label{14}
\end{equation}
\noindent As a result, the function $\alpha(x) =\varphi (x)+\beta$  depends on parameters $a=u_{rot}/u_0$, $\beta$ , the angles $\theta$ and $\varphi$.  The parameter $a$ depends on the ratio of the rotating velocity to the  velocity $u_0$, characterizing the thermal and turbulent velocities. This parameter can be accepted as the model parameters. Thus, we cannot estimate the rotation velocity $u_{rot}$ from the observations  in the explicit  form.

 The total radiation is characterized by the sum of the continuum radiation vector ${\bf I}_{cont}(\mu,\tau_{T})$ and the spectral line
radiation ${\bf I}_{res}(\tau,x,\mu)$.
  Further we consider the case, when the resonant atoms are located near the surface of the optically thick accretion disc and the radiation density of the continuum  is larger than that of the resonance line radiation.  It appears,  this is the  natural assumption about the location of the  resonant atom sources.  Besides, we consider that the continuum
radiation is the solution of the Milne problem (see Chandrasekhar 1960).

Below we shortly follow to general theory of resolvent matrices given in Silant'ev et al. 2015, 2017a.

\section{Solution of integral equation for ${\bf S}(\tau)$ using resolvent matrix}

 According to the standard theory of integral equations (see, for example, Smirnov 1964), the solution of  Eq.(7) can be presented in the following form  (see Silant'ev et al. 2015, 2017b):

\begin{equation}
{\bf  S}(\tau)={\bf g}(\tau)+\int\limits_0^{\infty}d\tau'\hat {R}(\tau,\tau')\,{\bf  g}(\tau'),
\label{15}
\end{equation}
\noindent where the resolvent matrix $\hat{R}(\tau,\tau')$ obeys the integral equation:
\begin{equation}
\hat{R}(\tau,\tau')=\hat{L}(|\tau-\tau'|)+\int\limits_0^{\infty}d\tau''\hat L(|\tau-\tau''|)\hat{R}(\tau'',\tau').
\label{16}
\end{equation}
\noindent The property $\hat L^T=\hat L$ gives rise to the relation $\hat{R}^T(\tau,\tau')=\hat{R}(\tau',\tau)$.  We see that the equation for $\hat{R}(\tau,0)$ follows from Eq.(16):

\begin{equation}
\hat{R}(\tau,0)=\hat{L}(\tau)+\int\limits_0^{\infty}d\tau'\hat L(|\tau-\tau'|)\hat{R}(\tau', 0).
\label{17}
\end{equation}
\noindent This equation is analogous to Eq.(7). It means that $\hat{R}(\tau,0)$ can be presented in the form:
\begin{equation}
\hat{R}(\tau,0)=\hat{L}(\tau)+\int\limits_0^{\infty}d\tau''\hat R(\tau,\tau')\hat{L}(\tau').
\label{18}
\end{equation}
\noindent The general theory (see Sobolev 1969, 
 Silant'ev et al. 2015) demonstrates that the resolvent $\hat{R}(\tau,\tau')$ can be calculated, if we know the matrices $\hat{R}(\tau,0)$ and $\hat{R}(0,\tau')$. This is seen directly from the expression for the double Laplace transform of $\hat{R}(\tau,\tau')$ with the parameters $b$ and $c$ :

\begin{equation}
\tilde{\tilde{\hat{R}}}(b,c)=\frac{1}{b+c}[\,\tilde{\hat{R}}(b,0)+\tilde{\hat{R}}(0,c)+\tilde{\hat{R}}(b,0)\tilde{\hat{R}}(0,c)].
\label{19}
\end{equation}

Taking the  Laplace transform of $\hat{R}(\tau,0)$  and using the relation (19), we can derive non-linear equation for $H$-matrix:

\begin{equation}
\hat{H}(z)=\hat{E} +\tilde{\hat{R}}\left(\frac{1}{z},0\right)\,\,\,,
\label{20}
\end{equation}
\noindent where $ \tilde{\hat{R}}(1/z,0)$ is the Laplace transform of  $\hat {R}(\tau,0)$ with the  parameter $1/z$.
 $\hat{E}$  is the unit matrix. This equation has the form:
\[
\hat{H}(z)=\hat{E} +
z\hat{H}(z)\int_{-\infty}^{\infty}dx'\varphi^2(x')\int_0^1d\mu'\times
\]
\begin{equation}
\frac{1}{\mu'+\alpha(x')z}\hat{H}^T\left(\frac{\mu'}{\alpha(x')}\right)\hat{\Psi}(\mu').
\label{21}
\end{equation}
\noindent The  $\hat{H}(z)$-matrix can be calculated, if we know the matrix $\hat{H}\left(\frac{\mu}{\alpha(x)}\right)$, which obeys the following non-linear equation:
\[
\hat{H}\left(\frac{\mu}{\alpha(x)}\right)=\hat{E} +\mu\hat{H}\left(\frac{\mu}{\alpha(x)}\right)\int_{-\infty}^{\infty}dx'\int_0^1d\mu'\times
\]
\begin{equation}
\frac{\varphi^2(x')}{\alpha(x)\mu'+\alpha(x')\mu}\hat{H}^T\left(\frac{\mu'}{\alpha(x')}\right)\hat\Psi(\mu').
\label{22}
\end{equation}
\noindent  Recall,  that the parameter $x$ is the dimensionless frequency (see Eq.(4)). The parameters $b,c,z$ are mathematical parameters, used in the integral equations theory.

 The $\hat H(z)$-matrix obeys the  linear equation. Following to the technique in  Silant'ev et al. 2015, we obtain:
\[
\left(\hat E-2z^2\int\limits_0^1d\mu\int\limits_{-\infty}^{\infty}dx\frac{\varphi^2(x) \alpha(x)\hat \Psi(\mu)}{(\alpha(x)z)^2-\mu^2}\right)H(z)=
\]
\begin{equation}
\left(\hat E-z\int\limits_0^1d\mu\int\limits_{-\infty}^{\infty}dx\frac{\varphi^2(x)\hat \Psi(\mu) \hat{H}\left(\frac{\mu}{\alpha(x)}\right)}{\alpha(x)z-\mu}\right).
\label{23}
\end{equation}

From Eq.(2) we obtain the expression for the emerging radiation:
\[
{\bf I}_{res}(0,x,\mu)=\]
\[
\varphi(x)\hat{A}(\mu)\int\limits_0^{\infty}\frac{d\tau }{\mu}\exp{\left (-\frac{\alpha(x)\tau}{\mu}\right )}{\bf S}_{res}(\tau)
\]
\begin{equation}
\equiv\frac{\varphi(x)}{\mu}\hat{A}(\mu)\tilde{{\bf  S}}_{res}\left (\frac{\alpha(x)}{\mu}\right ).
\label{24}
\end{equation}
\noindent The most simple formula arises for the exponential  source of non-polarized radiation:
\begin{equation}
{\bf g}(\tau)= \frac{1-\varepsilon}{2}\, s_0\exp{(-h\tau)}\left (\begin{array}{c}1\\0 \end{array}\right).
\label{25}
\end{equation}
\noindent Using the relation (19), we obtain:
\[
{\bf I}_{res}(0,x,\mu)=
\]
\begin{equation}
\frac{1-\varepsilon}{2}\,s_0\,\varphi(x)\hat{A}(\mu)\frac{\hat{H}\left(\mu/\alpha(x)\right)\hat H\left(1/h\right)\left (\begin{array}{c}1\\0 \end{array}\right)}{\alpha(x)+\mu\,h}.
\label{26}
\end{equation}
\noindent Note that $h=0$ corresponds to the homogeneous  source of non-polarized radiation. The results for sources of the type $s(\tau)\sim \tau, \tau^2, etc.$ can be also  obtain from  Eq.(26)(see Silant'ev et al. 2017a).

\section{Short comments on the  impossibility of  the Milne problem for resonance line}

 Recall, that the usual Milne problem describes the diffusion of the radiation through the semi-infinite non-absorbing atmosphere, where  the radiation flux is constant. For the spectral line diffusion the total flux ( introducing the integral over all frequences of a line ) is  not conserved   because the parameters $\varepsilon\neq 0$ and $\beta\neq 0$ (see Ivanov 1963).  Eq.(2) gives rise to the conservation of the total flux only for the case  $\varepsilon=0$ and $\beta =0$.

 The generalized Milne problem (see Silant'ev et al. 2017b) describes the radiation  diffusion through the semi-infinite atmosphere with the  absorption.  For this case the total flux decreases with the approaching to the surface. Nevertheless, we can obtain the intensity distribution $J(0,x,\mu)=I(0,x,\mu)/I(0,x,0)$ and the polarization degree $p(0,x,\mu)=Q(0,x,\mu)/I(0,x,\mu)$ of emerging radiation.  Below we consider the generalized Milne problem.

It is known (see Sobolev 1969), that the Milne problem in both cases corresponds to the solution of the radiative transfer equation without the free term ${\bf g}(\tau)$. In this case ${\bf S}(\tau)={\bf K}_{res}(\tau)$ and the equation (7) transforms to the  homogeneous equation for ${\bf K}_{res}(\tau)$:
\begin{equation}
{\bf K}_{res}(\tau)|_{hom}=\int\limits_0^{\infty} d\tau'\hat L(|\tau-\tau'|)\,{\bf K}_{res}(\tau')|_{hom}.
\label{27}
\end{equation} 
\noindent The Milne problem corresponds to non-zero solution of Eq.(27).

In this case   the vector ${\bf I}_{res}(0,x,\mu)$, describing the emerging radiation, has the form:
\[
{\bf I}_{res}(0,x,\mu)=
\]
\[
\varphi(x)\hat{A}(\mu)\int\limits_0^{\infty}\frac{d\tau }{\mu}\exp{\left (-\frac{\alpha(x)\tau}{\mu}\right )}{\bf K}_{res}(\tau)
\]
\begin{equation}
\equiv\frac{\varphi(x)}{\mu}\hat{A}(\mu)\tilde{{\bf  K}}_{res}\left (\frac{\alpha(x)}{\mu}\right ),
\label{28}
\end{equation}
\noindent  i.e. this expression is  proportional to the Laplace transform of ${\bf K}_{res}(\tau)$. 

 The values ${\bf I}_{res}(0,x,\mu)$ and $\tilde{{\bf  K}}_{res}\left (\frac{\alpha(x)}{\mu}\right )$ according to the general theory (see Silant'ev et al. 2017b) depend on ${\bf K}(0)$. The latter obeys the homogeneous equation:
\begin{equation}
\left(\hat E-\int\limits_0^1d\mu\int\limits_{-\infty}^{\infty}dx\frac{\varphi^2(x)\hat \Psi(\mu) \hat{H}\left(\frac{\mu}{\alpha(x)}\right)}{\alpha(x)-k\mu}\right){\bf K}(0)=0,
\label{29}
\end{equation}
\noindent which has non-zero solution, if the determinant  is equal to zero. This  equation allows to obtain the 
characteristic number $k$.  Taking $z=1/k$ in linear Eq.(23), we obtain that the  determinant is zero, if the characteristic equation is:

\[
\Delta(k)\equiv det (k)=
\] 
\begin{equation}
\left|\left(\hat{E}-2\int_{-\infty}^{\infty}dx\int_0^1 d\mu\frac{\alpha(x)\varphi^2(x)\hat{\Psi}(\mu)}{\alpha^2(x)- k^2\mu^2}\right)\right|=0.
\label{30}
\end{equation}
\noindent This equation does not depend on the $\hat H$-matrix.

Most simple case corresponds to  the scalar equation for the intensity and assume that the light scattering is isotropic. In this case $\Psi(\mu)=(1-\varepsilon)/2$ and the characteristic equation is:

\[
\Delta(k)\equiv det (k)=
\] 
\begin{equation}
\left|\left(1-(1-\varepsilon)\int_{-\infty}^{\infty}dx\int_0^1 d\mu\frac{\alpha(x)\varphi^2(x)}{\alpha^2(x)- k^2\mu^2}\right)\right|=0.
\label{31}
\end{equation}
\noindent Here we follow to the simple approach to solve the scalar Milne problem by Sobolev 1969. It is easy to obtain that for $k=0$ and $\beta=0$  the determinants of Eqs.(30) and (31) are zero, if $\varepsilon =0$. Recall, that $\varepsilon\neq 0$, i.e. the value $k=0$ does not exist as the solution of the characteristic equations.

  Following  to the technique of Silant'ev et al. 2017b, we derive the expression for ${\bf I}(0,x,\mu)$ in the form:
\begin{equation}
{\bf I}_{res}(0,x,\mu)=\varphi(x)\hat{A}(\mu)\hat{H}\left(\frac{\mu}{\alpha(x)}\right)\frac{{\bf K}_{res}(0)}{\alpha(x)-k\mu}.
\label{32}
\end{equation}
\noindent Here  $\alpha(x)=\varphi(x)+\beta$.  The positive intensity of the emerging radiation takes place for characteristic number $k<\beta$. Only for such $k$  the Milne problem for the resonance line has the solution. If $k>\beta$  the intensity of outgoing radiation (32) takes the negative value, i.e. we have the physically absurd result.   The solutions of characteristic equations (30) and (31) for parameter $k$  demonstrate  $k>\beta$, i. e.  the Milne problem for a spectral line is impossible, in contrast to the continuum radiation.

Let us discuss this result.
 Physical sense of the Milne problem means that there exists the conservation of the line  shape in the deep layers of an atmosphere, where the multiple scatterings of the light hold. The presence of the  non-resonant  atoms,  the  dust grains and non-zero parameter $\varepsilon$
destructs  the line shape and the Milne problem becomes  impossible. 

 The impossibility of the Milne problem for the spectral lines gives rise to the new interesting problem. We can to take the solution of the Milne problem  for the continuum radiation as the source function, i.e. take the radiation density as the sum  ${\bf K}_{res}(\tau)$ and  ${\bf K}_{cont}(\tau)$. Moreover, we assume that the resonant atoms locate in a thin layer near the surface of an accretion disc. In this case we can neglect the multiple scattering of the resonance radiation on the resonant atoms. This gives  ${\bf K}_{res}(\tau)<<  {\bf K}_{cont}(\tau)$.    Below we give the solution of this problem.

\section{The  Milne problem for  the continuum and  the calculation of the emerging resonance line intensity and polarization}

 Thus, we calculate the continuum radiation density ${\bf K}_{cont}(\tau_{T})$  as the solution of the Milne problem in the electron  atmosphere. Recall, that the radiative transfer equation for continuum radiation has the form:
\begin{equation}
\mu\frac{d{\bf I}_{cont}(\tau_T,\mu)}{d\tau_T}=-{\bf I}_{cont}(\tau_T,\mu)+\hat A(\mu){\bf S}(\tau_T),
\label{33}
\end{equation}
 where the vector ${\bf S}(\tau_T)$ is:

\[
{\bf S}(\tau_T)=\frac{1-\varepsilon}{2}\left (\begin{array}{c}s_I(\tau_T)\\s_Q(\tau_T)\end{array}\right) + {\bf K}_{cont}(\tau_T),
\]
\begin{equation}
{\bf K}_{cont}(\tau_T)=  \frac{1-\varepsilon}{2}\int\limits_{-1}^1\,d\mu\,\hat A^T(\mu){\bf I}_{cont}(\tau_T,\mu).
\label{34}
\end{equation}
\noindent The Thomson optical depth $d\tau_{T}=N_e\sigma_{T}dz$ is related with $d\tau$ by the following formula: $d\tau_{T}=\beta d\tau$.  This relation will to be used  in the further calculations, where we assume that the resonant atoms are distributed near the surface of the optically thick accretion disc. The latter condition gives rise the inequality  ${\bf K}_{cont}(\tau)>> {\bf K}_{res}(\tau)$ .

Recall, that we neglect by the radiation scattering on non-resonant atoms because the Thomson scattering cross-section $\sigma_T>>\sigma_{atom}$.

  In this situation the resonant radiation arises as a result of the scattering of the continuum radiation on the layer of the resonant atoms near the surface.
 Thus, we can  neglect ${\bf K}_{res}(\tau)$ compared with ${\bf K}_{cont}(\tau)$. Recall, that the vector ${\bf K}_{cont}(\tau)$  is determined by the Milne problem for continuum radiation in the electron  atmosphere. 

The scattering of this continuum radiation on the resonant atoms produces the emerging radiation ${\bf I}_{res}(0,x,\mu)$, describing by Eq.(28) , where  ${\bf K}_{res}(\alpha(x)/\mu) \to  {\bf K}_{cont}(\alpha(x)/\mu)$. As a result, the expression (28) transforms to:
\[
{\bf I}_{res}(0,x,\mu)=
\]
\[
\frac{\varphi(x)\hat{A}(\mu)}{\beta\mu}\int\limits_0^{\infty}d\tau_{T }\exp{\left (-\frac{\alpha(x)\tau_{T}}{\beta\mu}\right )}{\bf K}_{cont}(\tau_{T})
\]
\begin{equation}
\equiv\frac{\varphi(x)\hat{A}(\mu)}{\beta\mu}\tilde{{\bf  K}}_{cont}\left (\frac{\alpha(x)}{\beta\mu}\right ).
\label{35}
\end{equation}

 The vector ${\bf K}_{cont}(\tau)$ obeys Eq.(27), where the kernel $\hat L(|\tau-\tau'|)$ is equal to (see Silant'ev et al. 2017b):
\begin{equation}
\hat L(|\tau-\tau'|)=\int\limits_0^1\frac{d\mu}{\mu}\exp{\left(-\frac{|\tau-\tau'|}{\mu}\right)}\,\hat \Psi(\mu).
\label{36}
\end{equation}
\noindent  The value  $\hat \Psi(\mu)=(1-\varepsilon)/2)\hat A(\mu)^T \hat A(\mu)$, where $(1-\varepsilon)$ describes  the existence of the  pure absorption of a continuum radiation by the dust  grains and the scattering on non-resonant atoms.

The characteristic equation in this case is more simple:
\begin{equation}
\left|\left(\hat{E}-2\int_0^1 d\mu\frac{\hat{\Psi}(\mu)}{1- k^2\mu^2}\right)\right|=0.
\label{37}
\end{equation}
\noindent  The solution of this equation is $k\le1$ for every value of the absorption.   Note that the generalized Milne problem for continuum with  the parameter $\varepsilon \neq 0$ is considered in Silant'ev et al. 2017b. The  relation $k\le1$  means that  the Milne problem for continuum radiation always has the solution (see Eq.(41)).

Instead of Eq.(21) we have more simple expression:
\begin{equation}
\hat{H}(z)=\hat{E} +
z\hat{H}(z)\int_0^1d\mu'\frac{1}{\mu'+z}\hat{H}^T(\mu')\hat{\Psi}(\mu').
\label{38}
\end{equation}
According to the theory, presented in Silant'ev et al. 2017b, we obtain the formula for $\tilde{{\bf  K}}_{cont}\left (\alpha(x)/(\beta\mu)\right )$:
\begin{equation}
\tilde{{\bf  K}}_{cont}(\alpha(x)/(\beta\mu) )=\beta\mu\frac{\hat H(\beta\mu/\alpha(x)){\bf K}_{cont}(0)}{\alpha(x)-k\beta\mu}.
\label{39}
\end{equation}

  It is convenient to use  the new matrices:  $\hat D(\mu)=\hat A(\mu)\hat H(\mu)$ and  $\hat D(x,\mu)=\hat A(\mu)\hat H(\beta\mu/\alpha(x))$.
The matrix $\hat D(\mu)$ obeys  the following equation:

\[
\hat{D}(\mu)\equiv \left (\begin{array}{rr}a(\mu) ,\,\, b(\mu) \\ c(\mu),\,\,\,d(\mu) \end{array}\right)=
\]
\begin{equation}
\hat A(\mu)+
\frac{\mu(1-\varepsilon)}{2}\int_0^1d\mu'\frac{\hat D(\mu)\hat D^T(\mu')\hat A(\mu')}{\mu+\mu'}.
\label{40}
\end{equation}
\noindent  The emerging continuum radiation is described by the formula:
\begin{equation}
{\bf I}_{cont}(0,\mu)=\hat{D}(\mu)\frac{{\bf K}_{cont}(0)}{1-k\mu}.
\label{41}
\end{equation}
\noindent The denominator of Eq.(41) is positive  because the characteristic number $k\le1$ .

  The equation for matrix $\hat D(x,\mu)$ follows from Eq.(38), where $z=\beta\mu/\alpha(x)$:
\[
\hat D(x,\mu)=\hat A(\mu)+\frac{(1-\varepsilon)\mu}{2}\hat D(x,\mu)\times
\]
\begin{equation}
\int_0^1d\mu'\frac{1}{\mu+\mu'\alpha(x)/\beta}\hat D^T(\mu')\hat A(\mu').
\label{42}
\end{equation}
\noindent Eq.(42) can be presented in the form:
\[
\hat D(x,\mu)\left[ \hat E-\frac{(1-\varepsilon)\mu}{2}\times \right.
\]
\begin{equation}
\left.\int_0^1d\mu'\frac{1}{\mu+\mu'\alpha(x)/\beta}\hat D^T(\mu')\hat A(\mu')\right ]=\hat A(\mu),
\label{43}
\end{equation}
\noindent i. e. this equation can be readily solved, using the matrix $\hat D(\mu)$.  The technique of the numerical calculation of Eq.(40) is given in Silant'ev et al. 2017b.

As a result,  expression (35) reaches the form: 
\begin{equation}
{\bf I}_{res}(0,x,\mu)=\varphi(x)\hat{D}(x,\mu)\frac{{\bf K}_{cont}(0)}{\alpha(x)- k\beta\mu}.
\label{44}
\end{equation}
\noindent  Note that the denominator of Eq.(44) is positive ($\alpha(x)-\beta k \mu=(\varphi(x)+\beta(1-k\mu))>0)$. 

 The homogeneous equation (27) (for ${\bf K}_{cont}(\tau)$) allows us to obtain the ratio $K_Q(0)/K_I(0)=t$. We take $K_I(0)=1$ and $K_Q(0)=t$.  As a result, from  Eq.(44)  we can  obtain the angular distribution
 $J_{res}(x,\mu)\equiv I_{res}(0,x,\mu)/I_{res}(0,0,\mu)$ and the degree of polarization $p_{res}(x,\mu)\equiv Q_{res}(0,x,\mu)/I_{res}(0,x,\mu)$. The value  $J_{res}(x,\mu)$ describes the shape of the spectral line.  The negative polarization degree $p_{res}$ denotes that the wave electric field oscillations are  perpendicular to the plane ${\bf (nN)}$.
Recall, that  the characteristic number $k$ obeys  Eq.(37) for continuum radiation.

Below we consider the free electron conservative atmosphere, where $\varepsilon=0$ and $k=0$. In this case the parameter $t= -0.106280$. 

The value $\alpha(x)/\beta=\varphi(x)/\beta+1$ tends to $1$ for $x\gg 1$. In this case $ \hat D(x,\mu) \to\hat D(\mu)$. Asymptotically Eq.(44) for $x>>1$ tends to:
\begin{equation}
{\bf I}_{res}(0,x,\mu)\to\frac{\varphi(x)}{\beta}{\bf I}_{cont}(0,\mu).
\label{45}
\end{equation}
\noindent In particular,  Eq.(45) gives rise to asymptotical relation $p_{res}(x,\mu)\to p_{cont}(\mu)$, i.e. the polarization degree  of the resonance line in the  far wings coincides with that of the continuum radiation.  Our numerical calculations, using the technique Silant'ev et al. 2017b (see Figs.4-9), confirm this.
 Physically this is natural because the far wings coincide with the continuum radiation.

 Formula (44) corresponds to the accretion disc with the  vertical temperatute the highest in the central plane of an accretion disc, i.e. this allows us to use the Milne problem theory for continuum radiation.  Such temperature corresponds to various models of accretion discs (see, for example, Shakura \& Sunyaev 1973)).  The diffusion of continuum radiation from the central plane to the surface gives rise to the radiation density vector ${\bf K}_{cont}(\tau_{T})$.  This continuum radiation, scattered on the resonant atoms,  produces the resonance line emission.
 
Note that most important assumptions in our theory are ${\bf K}_{cont}>>{\bf K}_{res}$ and that the high temperature exists in the central plane of the optically thick  accretion disc.  These assumptions allow us  to  neglect by the multiple scatterings of the resonance line radiation. 

\section{Results of calculations}

Now we discuss the results of the numerical calculations of Eq.(35) for the angular distribution $J_{res}(x,\mu)=I_{res}(0,x,\mu)/I_{res}(0,0,\mu)$ and the polarization degree $p_{res}(x,\mu)=Q_{res}(0,x,\mu)/I_{res}(0,x,\mu)$ of the resonance radiation emerging from the optically thick accretion disc, consisting of  free electrons and the thin layer of the resonant atoms near the surface. The technique of numerical calculations is given in Silant'ev et al 2017b . The results of the calculations are given in Figures 4-9 and Tables 1-3.

Below we will show the coincidence of our effects with those in a number of  sources given  in the  spectropolarimetric atlas of Seyfert 1 galaxies (see Smith et al. 2002). The particular models, presented in Smith et al. 2004, 2005, we do not discuss.

  Firstly we note that the polarization  $p_{res}(x,\mu)$ is negative.  This corresponds to the wave electric field oscillations perpendicular to the plane $({\bf nN})$, i.e. as in known solution of the Milne problem for continuum radiation (see Chandrasekhar 1960). This is natural because  the scattering matrix $\hat A(\mu)\hat A^T(\mu')$ in our transfer equation (2) is similar to that  in the radiative transfer equation (33) for continuum radiation in an electron atmosphere.

According to Chandrasekhar's solution,  the angular distribution $J(\mu)=I_{cont}(0,\mu)/I_{cont}(0,0)$ has the maximum value $3.063$ at $\mu=1$ and the maximum polarization degree $11.713\%$ at $\mu=0 (\theta=90^{\circ})$. Due to the axial symmetry, the polarization at $\mu=1(\theta=0^{\circ})$ is equal to zero both for the continuum and resonance radiations.

 For the resonance line the polarization degree $11.713\%$  holds for all frequencies $x$ at $\mu=0$.  For $\mu=0$ the polarization arises due to the last scattering of continuum radiation both on free electrons  and  on resonant atoms.   For  angles $\theta \neq  90^{\circ}$  the polarization degree depends on the angle  $\theta$ and  the parameters  $\beta$ and $a=u_{rot}/u_0$.    Recall, that in the far wings of the resonance line the polarization degree $p_{res}(x,\mu)\to p_{cont}(\mu)$ for all values of $\mu$ (see Eq.(45)).

First of all,  we consider the case of non-rotating accretion disc ($u_{rot}=0$).  It is clear that  in this case all parts of the non-rotating disc are equivalent one to another. Therefore  we give the results only in Figs. 4 and 5.

  In Tables 1, 2 and 3 we consider  the angular distribution $J_{res}(x,\mu)$ and the polarization degree  $p_{res}(x,\mu)\%$ for $\theta=0^{\circ}, 30^{\circ}, 45^{\circ}, 60^{\circ}, 90^{\circ}$.   We took $\beta=0.5, 1, 2$ and $a=0$. These Tables  correspond to non-rotating accretion disc. Due to the axial symmetry the right and left wings of the resonance line have similar form. We present only the right wing. For large $x> 2-3$ the $I_{res}(x,\mu)\to 0$ and $p_{res}(x,\mu)\to p_{cont}(\mu)$ in accordance with Eq.(45). These Tables and analogous ones for rotating disc can be used for the estimation of the inclination angle  $i\equiv \theta$ of an accretion disc. This angle is very important parameter of the accretion disc.

\begin{figure*}[h!]
\includegraphics [width=16cm, height=8cm]
{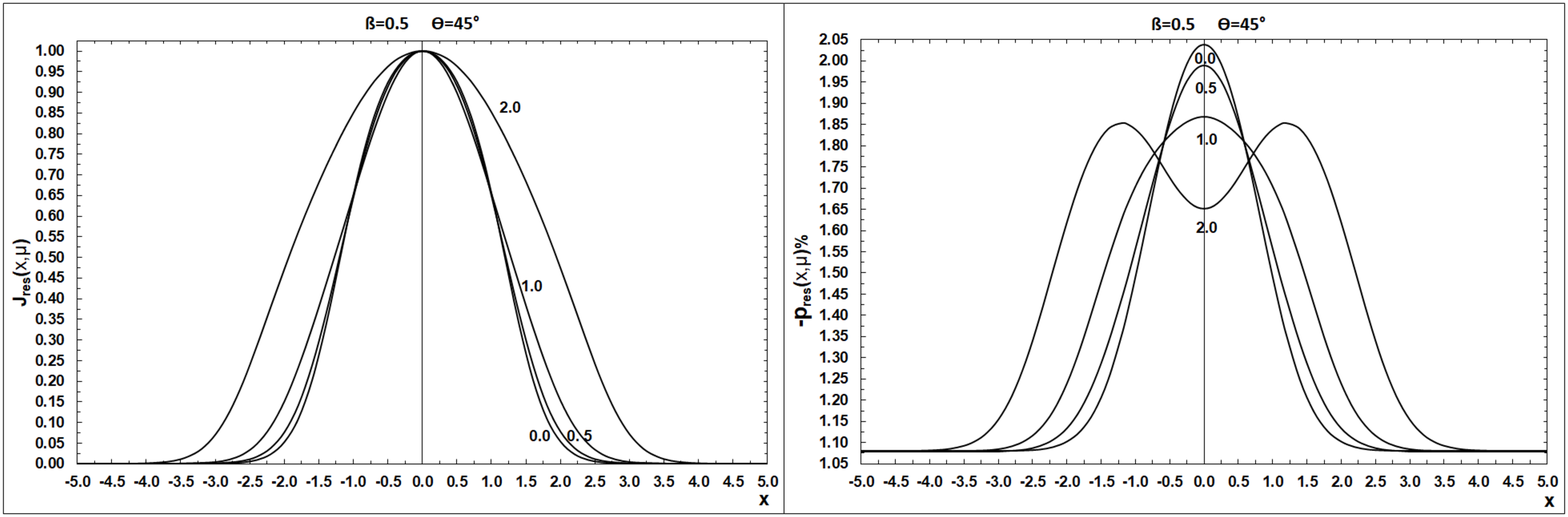}
\caption{The result of numerical calculations of Eq.(35) gives the following angular distribution $J_{res}(x,\mu)$  and polarization degree $p_{res}(x,\mu)$ \% for $\beta=0.5$ and $\theta=45^{\circ}$. 
The homogeneous sources  are distributed  axially symmetric around the center of accretion disc. The numbers denote the parameter $a=u_{rot}/u_0 $. }
\label{a}
\end{figure*}

\begin{figure*}[h!]
\includegraphics [width=16cm, height=8cm]
{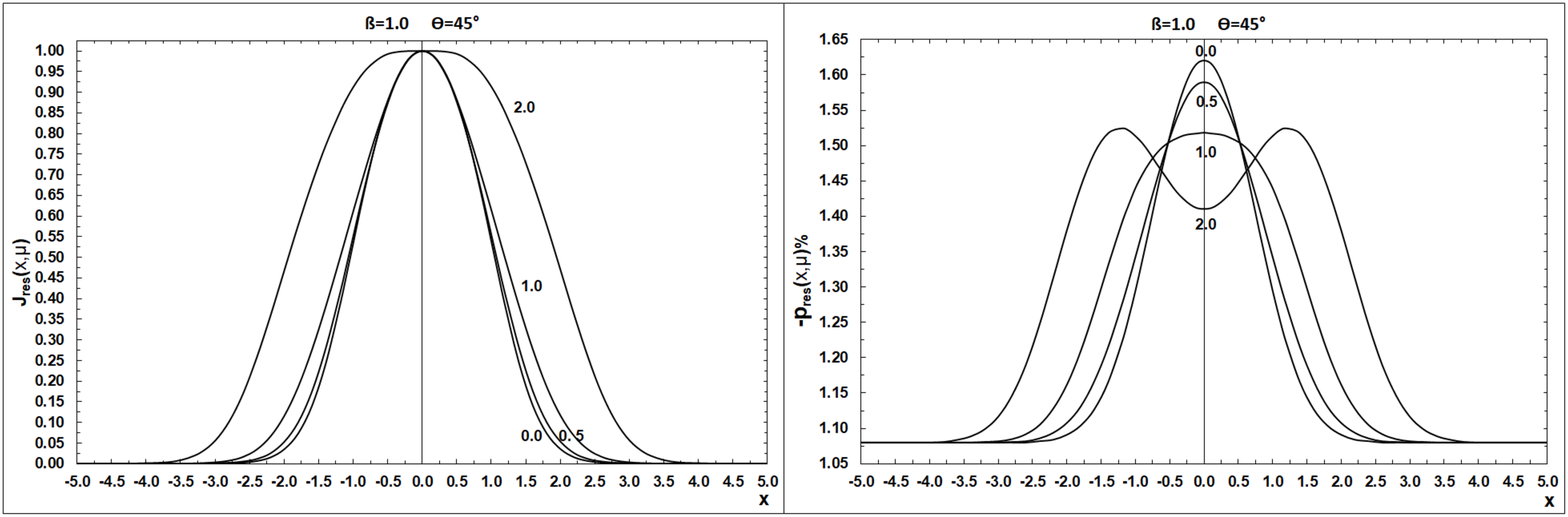}
\caption{The result of numerical calculations of  Eq.(35) gives the following  angular  distribution $J_{res}(x,\mu)$  and polarization degree $p_{res}(x,\mu)$ \% for $\beta=1.0$ and $\theta=45^{\circ}$. The homogeneous sources  are distributed  axially symmetric around the center of accretion disc. The numbers denote the parameter $a=u_{rot}/u_0 $. }    
\label{Silan_Fig5.eps}
\end{figure*}

\begin{figure*}[h!]
\includegraphics [width=16cm, height=8cm]
{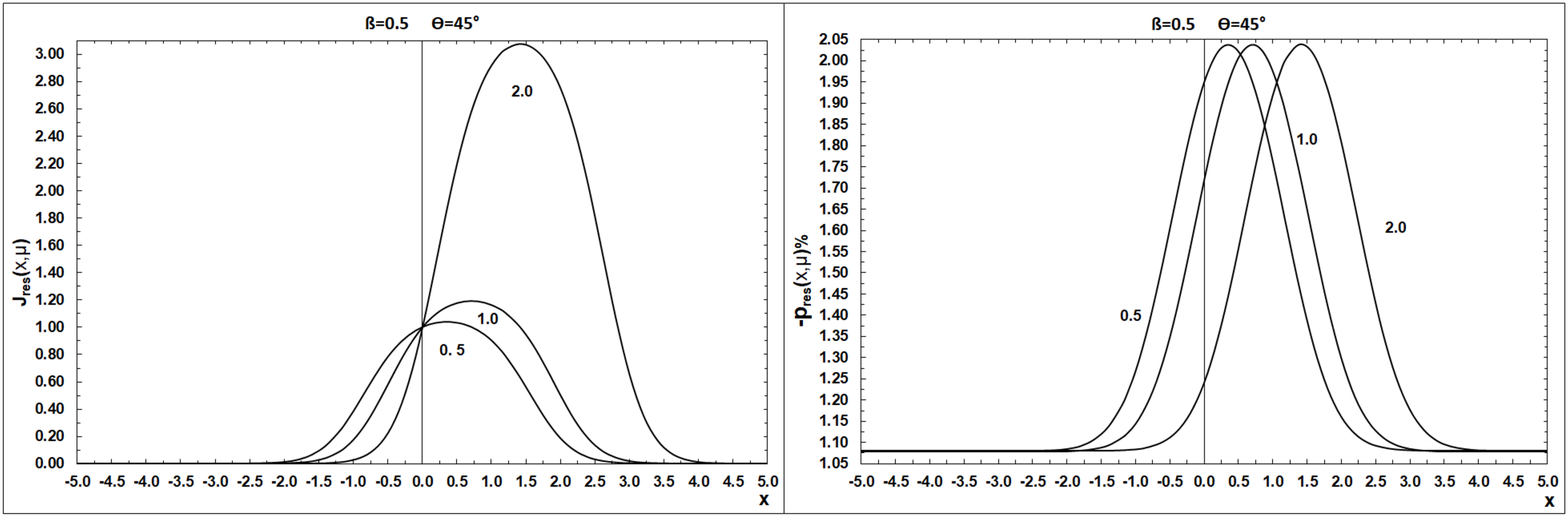}
\caption{The result of numerical calculations of  Eq.(35) gives the following  angular distribution $J_{res}(x,\mu)$  and polarization degree $p_{res}(x,\mu)$ \% for $\beta=0.5$ and $\theta=45^{\circ}$  for spot-like source with the maximum Doppler shift ($\varphi=90^{\circ}$).  The numbers denote the parameter $a=u_{rot}/u_0 $. }
\label{Silan_Fig6.eps}
\end{figure*}

 Now we consider the rotating discs.
In Figs. 4 and 5 we present the  $J_{res}(x,\mu)$ and $p_{res}(x,\mu)$ for parameters $\beta=0.5$ and $\beta=1$ for rotating accretion discs with parameters $a=0, 0.5, 1, 2$,  when the resonant atoms are distributed on the orbit axially symmetric around the center of the accretion disc (see Fig.1.)  In these cases the  $J_{res}(x,\mu)$ and $p_{res}(x,\mu)$ are similar in the right and left wings of the emitting line. The numbers near curves denote the values of parameter $a=u_{rot}/u_0$.  Note that we take the angle $\theta=45^{\circ}$. The parameter $a=0$ corresponds to non-rotating disc.

Figs. 4 and 5 show that the shapes $J_{res}(x,\mu)$ have one peak-like form. They are  practically identical for $a=0,0.5$ and $1$. For the case $a=2$ the angular distributions  $J_{res}(x,\mu)$ differs essentially from those for $a=0,0.5,1$.  Beginning from $\beta=1$ the line shape at $a=2$ acquires the plane  form  in the interval $x\simeq (-0.5, 0.5)$.

 For polarization degrees the difference between 
$a=0$ and 0.5  is not profound.  Note that for $a=2$ the value $p_{res}(x,\mu)$ acquires two peaks at $x\simeq \pm 1.4$. It appears, this is the consequence that at $a=2$ the right and left wings are far from one to another. Clear that two peak  effect can be more profound for $a>2$.

  Two-peak  behaviour of the polarization degree is observed in H$\alpha$-line for the sources Mrk-6, Mrk-290 , Mrk-590, Mrk-985  and some other (see Smith et al. 2002). Recall, that the H$\alpha$-line has $\lambda_0=0.6563{\AA}$ in the  laboratory frame of reference. The observed H$\alpha$ - lines have cosmological  red shift, various  for different sources. 

It is of interest to consider the rotating accretion disc with the one spot-like distribution of the resonant atoms (see Fig.2, 6, 7). We take the spot with maximum  rotation, corresponding to $\varphi=90^{\circ}$ (see Eqs. (13,14)).  Figs. 6 and 7 present  $J_{res}(x,\mu)$ and  $p_{res}(x,\mu)$ for this spot-like source of radiating atoms. This  is not axially symmetric problem relative $x=0$.

  As in Figs. 4 and 5, we take $\theta=45^{\circ}$. The intensity of radiation in the right wing has the factor to $\exp{[-(x-a\sin\theta)^2]}$ and in the left wing has the factor  $\exp{[-(x+a\sin\theta)^2]}$. The maximum values of   $J_{res}(x,\mu)$ and $p_{res}(x,\mu)$  are independent of parameters $\beta$ and for $a=0.5, 1, 2$ hold at $x\simeq 0. 35,0.7 , 1.4$, correspondingly. The polarization degrees  $p_{res}(x,\mu)$ have the similar profiles (with the $p_{max}=1.62\%$ ) as for the case $a=0$, but are shifted from $x=0$ to $x\simeq 0.35, 7, 1.4$. It appears, such behaviour confirms that the emerging radiation holds polarization as a result of the last scattering of radiation before escape from the accretion disc. Recall, that polarization in wings corresponds  to values for continuum radiation $p_{cont}(\mu)$ (see Eq.(45)).

  The shift of the resonance line frequency for the intensity and polarization degree is proportional to the parameter  $a=u_{rot}/u_0$.  The one peak polarization degree is observed in some cases, for example,  in the source Fairall 51 (see Smith et al. 2002).

In Figs. 8 and 9 we present the  $J_{res}(x,\mu)$  and   $p_{res}(x,\mu)$ for the  two spot-like equal sources with $\beta=1$, which are located oppositely one to another ($\varphi =90^{\circ}$ and $270^{\circ}$, respectively). This is the axially symmetric case. Here we take  $\theta = 45^{\circ}$ and $30^{\circ}$.  The existence of two sources gives rise to more complex pictures than in Figs. 4 and 5. So, for $a=2$ and $\theta=45^{\circ}$  two peak-like shapes appear for $J_{res}(x,\mu)$(the maximums at $x\simeq\pm1.4$). For $\theta=30^{\circ}$ the  shape of line is practically plane in the  interval  $x\simeq(- 0.5, 0.5)$.

 It is interesting that frequently the  shape of the observed resonance  line (see Smith et al. 2002) is slighly different at the right and the left parts  of the peak . This effect can arise if the sources of the resonant atoms  in the left and right parts  have slightly different numbers of the resonant atoms.   The degree of polarization increases with
increase of $\theta $. This is similar to the case of the Milne problem for continuum radiation (see Chandrasekhar 1960).

In this paper we did not obtain the opposite  position angles for the left and the right wings. This effect is frequently observed (see Smith et al.2002). To obtain this effect  we need to have additional  scattering of resonance radiation from rotating disc on some cloud (see Smith et al. 2005). The another explanation is connected with the existence of magnetic field in the rotating  accretion disc. In this case the right and left wings affected by opposite Faraday rotations (see Silant'ev et al. 2013).  

\begin{figure*}[h!]
\includegraphics [width=16cm, height=8cm]
{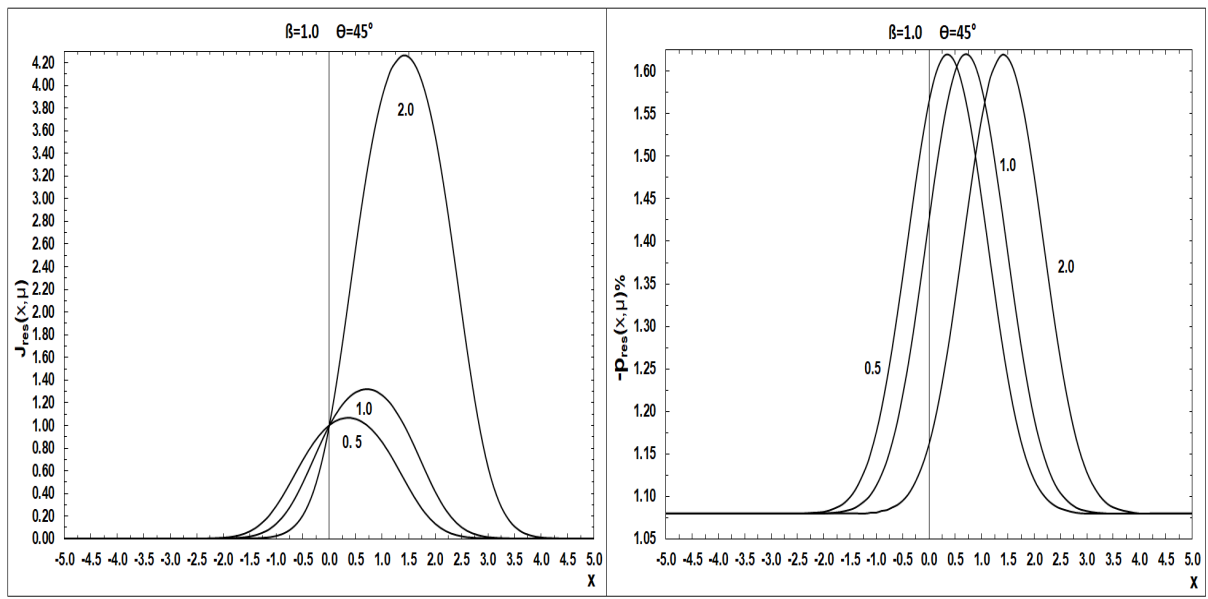}
\caption{The result of numerical calculations of  Eq.(35) gives the following  angular distribution $J_{res}(x,\mu)$  and polarization degree $p_{res}(x,\mu)$ \% for $\beta=1.0$ and $\theta=45^{\circ}$  for spot-like source  with the maximum Doppler shift ($\varphi=90^{\circ}$).  The numbers denote the  parameter $a=u_{rot}/u_0$.  }
\label{Silan_Fig7.eps}
\end{figure*}

\begin{figure*}[h!]
\includegraphics [width=16cm, height=8cm]
{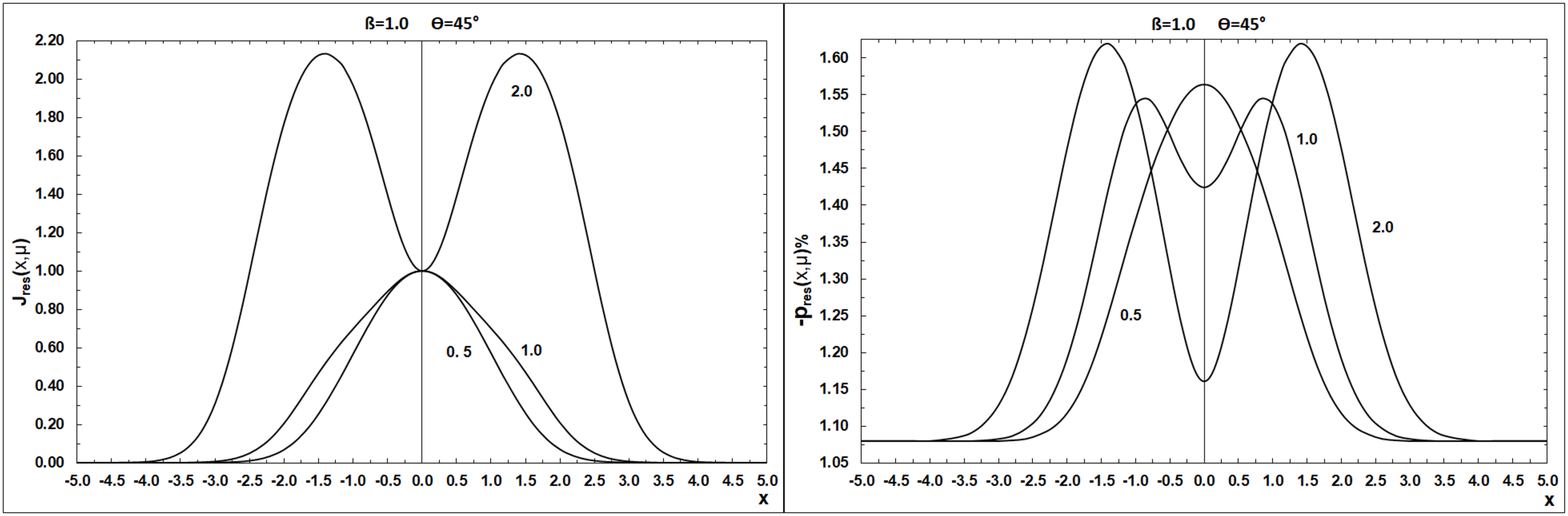}
\caption{The result of numerical calculations of  Eq.(35) gives the following  angular distribution $J_{res}(x,\mu)$  and polarization degree $p_{res}(x,\mu)$ \% for $\beta=1.0$ and $\theta=45^{\circ}$  for two equal spot-like sources  located at $\varphi =90^{\circ}$ and $\varphi= 270^{\circ}$. The numbers denote the  parameter $a=u_{rot}/u_0$.  }
\label{Silan_Fig8.eps}
\end{figure*}

\begin{figure*}[h!]
\includegraphics [width=16cm, height=8cm]
{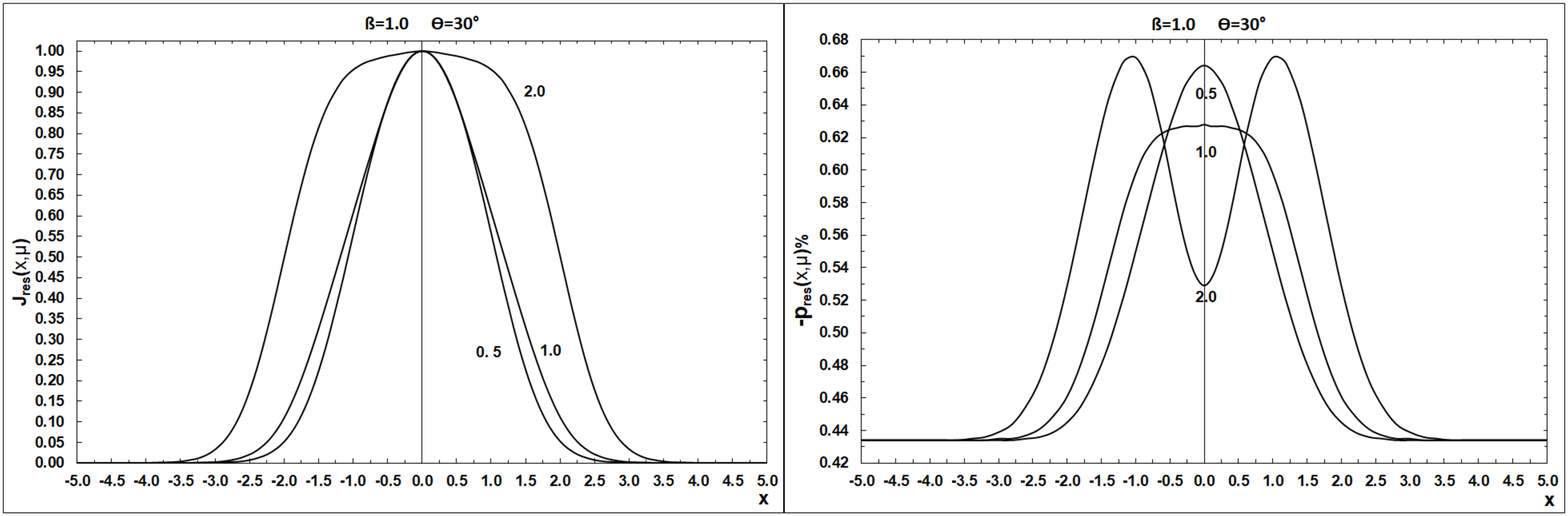}
\caption{The result of numerical calculations of  Eq.(35) gives the following  angular  distribution $J_{res}(x,\mu)$  and polarization degree $p_{res}(x,\mu)$ \% for $\beta=1.0$ and $\theta=30^{\circ}$  for two equal spot-like sources  located at $\varphi =90^{\circ}$ and $\varphi= 270^{\circ}$. The numbers denote the  parameter $a=u_{rot}/u_0$.  }
\label{Silan_Fig9.eps}
\end{figure*}

\section{\bf Conclusion}

 In this paper we consider the emerging of the resonance line radiation from the optically thick accretion disc, consisting of free electrones and resonant atoms. The accretion disc is assumed to be rotating.  We assume  that  the resonant atoms are located in thin layer below the surface of an accretion disc. We also assume that the density of  continuum radiation is due to the Milne problem  and is much greater than that for the resonant radiation.  Recall, that the Milne problem describes the emerging radiation from semi-infinite atmosphere,  if the sources of a radiation are located far below the surface. We consider that the source of continuum radiation located in the cental plane of an accretion disc,  where the temperature is the highest.  For continuum radiation the Milne problem holds always.  We shortly demonstrate that the Milne problem for the resonance radiation is impossible. In our case the resonance emission arises as a result of the scattering of continuum radiation on the thin layer of resonant atoms.

Three cases of the resonant atoms sources are considered. The first case corresponds to continuous  distribution of the  resonant atoms along the circular orbit (see Fig.1).  The second case corresponds to one spot-like  source {see Fig. 2). The third  case corresponds to two spot-like sources located  one oppositely to another(see Fig.3).  In the first and the third  cases the right and left wings of emerging line are similar. In the second case the shape of the  line radiation is shifted as compared with non-rotating accretion disc.  The shape of the emerging line intensity and polarization  depends significantly on the  ratio of the rotation velocity value to the velocity, characrerizing the Doppler width. It also  depends on the ratio of the electron number density to the number density of resonant atoms. The results of the calculations characterize the different  observational effects  of H$\alpha$ radiation in the accretion discs and can be used for estimations of parameters mentioned above. They also can be used for estimation of the inclination angle of an accretion disc.   The calculated values are compared with the shapes of the intensity and polarization  of  H$\alpha$ resonance lines observed in Seyfert-1 AGNs (see Smith et al 2002).

{\bf Acknowledgements.} 
This research was supported by the Program of Presidium of Russian Academy of Sciences N 12. 
Authors are very grateful to  referees for very useful remarks and advices.

\begin{table}[h]
\caption { \small The angular distribution $J(x,\mu)$ and degree of polarization  $-p_{res}(x,\mu)=- Q_{res}(x,\mu)/I_{res}(x,\mu)$ in \% for $\beta=0.5, a=0$  and
 different angles  $\vartheta^{\circ}$  for non-rotating accretion disc.  Note that $p_{res}(x,1)\equiv 0$ and $p_{res}(x,0)\equiv -11.713\%$.}
\scriptsize
\begin{tabular}{ r | p{0.5cm}p{0.5cm}p{0.5cm}p{0.5cm}p{0.5cm}p{0.5cm}p{0.4cm}p{0.3cm}}
\hline
\noalign{\smallskip}
$x$ & $J(0^{\circ})$ & $J(30^{\circ})$ & -$p(30^{\circ})$ & $J(45^{\circ})$ &-$p(45^{\circ})$ & $J(60^{\circ})$ &-$p(60^{\circ})$ & $J(90^{\circ})$ \\
\hline
\noalign{\smallskip}
0     & 1         & 1        & 0.872 & 1        &  2.038 & 1         & 3.833  &  1     \\
0.1 &  0.998 & 0.997 & 0.869 & 0.997 & 2.031 & 0.997  & 3.822  & 0.995  \\
0.2 &  0.990 & 0.989 & 0.859 & 0.989 & 2.009 & 0.987  & 3.788  & 0.981   \\
0.3 &  0.977 & 0.976 & 0.842 & 0.974 & 1.973 & 0.971  & 3.732  & 0.958   \\
0.4 &  0.958 & 0.956 & 0.819 & 0.953 & 1.925 & 0.948  & 3.657  & 0.925   \\
0.5 &  0.932 & 0.929 & 0.791 & 0.925 & 1.866 & 0.918  & 3.564  & 0.882   \\
0.6 &  0.898 & 0.894 & 0.759 & 0.888 & 1.798 & 0.879  & 3.456  & 0.831  \\
0.7 &  0.855 & 0.850 & 0.725 & 0.843 & 1.724 &  0.831 & 3.338  & 0.771  \\
0.8 &  0.803 & 0.797 & 0.689 & 0.788 & 1.646 & 0.774  & 3.212  & 0.704    \\
0.9 &  0.741 & 0.734 & 0.653 & 0.725 & 1.567 & 0.709  & 3.084  & 0.630  \\
1.0 &  0.671 & 0.663 & 0.618 & 0.653 & 1.491 & 0.636  & 2.958  & 0.553     \\
1.1 &  0.593 & 0.586 & 0.585 & 0.575 & 1.419 & 0.558  & 2.837  & 0.475      \\
1.2 &  0.511 & 0.504 & 0.556 & 0.494 & 1.354 & 0.478  & 2.727  & 0.398   \\
1.4 &  0.349 & 0.344 & 0.508 & 0.336 & 1.247 & 0.322  & 2.544  & 0.259    \\
1.6 &  0.212 & 0.208 & 0.475 & 0.203 & 1.173 & 0.194  & 2.416  & 0.151       \\
1.8 &  0.115 & 0.112 & 0.455 & 0.109 & 1.127 & 0.104  & 2.336  & 0.080  \\
2.0 &  0.056 & 0.054 & 0.444 & 0.053 & 1.102 & 0.050  & 2.292  & 0.038  \\
2.2 &  0.024 & 0.024 & 0.438 & 0.023 & 1.089 & 0.022  & 2.269  & 0.017    \\
2.4 &  0.010 & 0.010 & 0.436 & 0.009 & 1.083 & 0.009  & 2.259  & 0.007    \\
2.6 &  0.004 & 0.004 & 0.435 & 0.003 & 1.081 & 0.003  & 2.254  & 0.002   \\
2.8 &  0.001 & 0.001 & 0.434 & 0.001 & 1.080 & 0.001  & 2.253  & 0.001   \\
3.0 &  0.000 & 0.000 & 0.434 & 0.000 & 1.080 & 0.000  & 2.252  & 0.000  \\
\hline 
\end{tabular}
\end{table}

\begin{table}[h]
\caption {\small  The angular distribution $J_{res}(x,\mu)$ and degree of polarization $- p_{res}(x,\mu)=- Q_{res}(x,\mu)/I_{res}(x,\mu)$ in \% for $\beta=1, a=0$ and
 different angles  $\vartheta^{\circ}$ for non-rotating accretion disc.  Note that  $p_{res}(x,1)\equiv 0$ and $p_{res}(x,0)\equiv -11.713\%$.}
\scriptsize
\begin{tabular}{ r | p{0.5cm}p{0.5cm}p{0.5cm}p{0.5cm}p{0.5cm}p{0.5cm}p{0.5cm}p{0.4cm}}
\hline
\noalign{\smallskip}
$x$ & $J(0^{\circ})$ & $J(30^{\circ})$ & -$p(30^{\circ})$ & $J(45^{\circ})$   &-$p(45^{\circ})$  & $J(60^{\circ})$ &-$p(60^{\circ})$& $J(90^{\circ})$\\
\hline
\noalign{\smallskip}
0     & 1         & 1        & 0.677 & 1        & 1.620 & 1         & 3.170   & 1      \\
0.1 &  0.995 & 0.995 & 0.675 & 0.995 & 1.616 & 0.995  & 3.163  & 0.994  \\
0.2 &  0.981 & 0.981 & 0.668 & 0.981 & 1.602 & 0.980  & 3.140  & 0.975   \\
0.3 &  0.958 & 0.957 & 0.658 & 0.956 & 1.579 & 0.954  & 3.103  & 0.943   \\
0.4 &  0.925 & 0.924 & 0.644 & 0.922 & 1.549 & 0.918  & 3.054  & 0.900   \\
0.5 &  0.883 & 0.881 & 0.628 & 0.878 & 1.513 & 0.873  & 2.993  & 0.846   \\
0.6 &  0.830 & 0.828 & 0.609 & 0.824 & 1.472 & 0.818  & 2.925  & 0.783  \\
0.7 &  0.769 & 0.766 & 0.589 & 0.762 & 1.428 & 0.754  & 2.852  & 0.712  \\
0.8 &  0.700 & 0.697 & 0.568 & 0.691 & 1.382 & 0.683  & 2.776  & 0.636    \\
0.9 &  0.625 & 0.621 & 0.548 & 0.616 & 1.338 & 0.606  & 2.700  & 0.556 \\
1.0 &  0.546 & 0.542 & 0.529 & 0.536 & 1.295 & 0.527  & 2.628  & 0.477     \\
1.1 &  0.466 & 0.462 & 0.512 & 0.457 & 1.256 & 0.448  & 2.560  & 0.399      \\
1.2 &  0.388 & 0.385 & 0.496 & 0.380 & 1.221 & 0.371  & 2.500  & 0.327   \\
1.4 &  0.249 & 0.247 & 0.471 & 0.243 & 1.165 & 0.237  & 2.402  & 0.204    \\
1.6 &  0.144 & 0.143 & 0.455 & 0.140 & 1.127 & 0.136  & 2.335  & 0.116       \\
1.8 &  0.076 & 0.075 & 0.444 & 0.073 & 1.104 & 0.071  & 2.294  & 0.060   \\
2.0 &  0.036 & 0.036 & 0.439 & 0.035 & 1.091 & 0.034  & 2.272  & 0.028  \\
2.2 &  0.016 & 0.015 & 0.436 & 0.015 & 1.084 & 0.015  & 2.260  & 0.012    \\
2.4 &  0.006 & 0.006 & 0.435 & 0.006 & 1.082 & 0.006  & 2.255  & 0.005    \\
2.6 &  0.002 & 0.002 & 0.434 & 0.002 & 1.080 & 0.002  & 2.253  & 0.002       \\
2.8 &  0.001 & 0.001 & 0.434 & 0.001 & 1.080 & 0.001  & 2.252  & 0.001      \\
3.0 &  0.000 & 0.000 & 0.434 & 0.000 & 1.080 & 0.000  & 2.252  & 0.007  \\
\hline 
\end{tabular}
\end{table}

\begin{table}[h]
\caption {\small  The angular distribution $J_{res}(x,\mu)$ and degree of polarization $- p_{res}(x,\mu)=- Q_{res}(x,\mu)/I_{res}(x,\mu)$ in \% for $\beta=2, a=0$ and
 different angles  $\vartheta^{\circ}$ for non-rotating accretion disc. Note that  $p_{res}(x,1)\equiv 0$ and $p_{res}(x,0)\equiv -11.713\%$.}
\scriptsize
\begin{tabular}{ r |p{0.5cm}p{0.5cm}p{0.5cm}p{0.5cm}p{0.5cm}p{0.5cm}p{0.5cm}p{0.4cm}}
\hline
\noalign{\smallskip}
$x$ & $J(0^{\circ})$ & $J(30^{\circ})$ & -$p(30^{\circ})$ & $J(45^{\circ})$   &-$p(45^{\circ})$  & $J(60^{\circ})$ &-$p(60^{\circ})$& $J(90^{\circ})$\\
\hline
\noalign{\smallskip}
0       & 1       & 1        & 0.562 & 1        &  1.368 & 1       & 2.751   & 1        \\
0.1 &  0.993 & 0.993 & 0.561 & 0.993 & 1.365 & 0.993 & 2.746  & 0.992 \\
0.2 &  0.974 & 0.973 & 0.557 & 0.973 & 1.357 & 0.973 & 2.733  & 0.969   \\
0.3 &  0.941 & 0.941 & 0.551 & 0.940 & 1.344 & 0.939 & 2.711  & 0.932  \\
0.4 &  0.897 & 0.896 & 0.544 & 0.895 & 1.327 & 0.893  & 2.683  & 0.881   \\
0.5 &  0.841 & 0.840 & 0.535 & 0.838 & 1.307 & 0.836  & 2.648  & 0.819  \\
0.6 &  0.776 & 0.775 & 0.525 & 0.773 & 1.285 & 0.769  & 2.609  & 0.747  \\
0.7 &  0.703 & 0.702 & 0.514 & 0.699 & 1.261 & 0.695  & 2.568  &  0.670  \\
0.8 &  0.625 & 0.624 & 0.503 & 0.621 & 1.236 & 0.616  & 2.526  &  0.589    \\
0.9 &  0.545 & 0.543 & 0.492 & 0.540 & 1.212 & 0.535  & 2.485  & 0.507  \\
1.0 &  0.465 & 0.463 & 0.483 & 0.460 & 1.190 & 0.455  & 2.446  & 0.427     \\
1.1 &  0.387 & 0.386 & 0.473 & 0.383 & 1.169 & 0.379 & 2.410  & 0.353   \\
1.2 &  0.316 & 0.314 & 0.465 & 0.312 & 1.151 & 0.308 & 2.378  & 0.285   \\
1.4 &  0.196 & 0.194 & 0.453 & 0.193 & 1.122 & 0.190 & 2.328  & 0.174    \\
1.6 &  0.110 & 0.110 & 0.444 & 0.109 & 1.103 & 0.107 & 2.294  & 0.097   \\
1.8 &  0.057 & 0.056 & 0.439 & 0.056 & 1.092 & 0.055 & 2.273  & 0.050   \\
2.0 &  0.027 & 0.027 & 0.436 & 0.026 & 1.085 & 0.026 & 2.262  & 0.023  \\
2.2 &  0.012 & 0.012 & 0.435 & 0.011 & 1.082 & 0.011 & 2.256 & 0.010   \\
2.4 &  0.005 & 0.005 & 0.434 & 0.005 & 1.081 & 0.004 & 2.254  & 0.004    \\
2.6 &  0.002 & 0.002 & 0.434 & 0.002 & 1.080 & 0.002 & 2.252  & 0.001    \\
2.8 &  0.001 & 0.001 & 0.434 & 0.001 & 1.080 & 0.001 & 2.252  & 0.001    \\
3.0 &  0.000 & 0.000 & 0.434 & 0.000 & 1.080 & 0.000  & 2.252 & 0.000  \\
\hline 
\end{tabular}
\end{table}

\end{document}